\documentclass[nohyperref]{article}
\usepackage[accepted]{icml2022}

\usepackage{aas_macros}
\usepackage{showyourwork}

\usepackage{microtype}
\usepackage{graphicx}
\usepackage{subfigure}
\usepackage{natbib}
\usepackage{booktabs} % for professional tables

\usepackage{hyperref}
\usepackage{xspace} % Space only when needed.
\usepackage{xcolor}
\definecolor{rb4}{HTML}{27408B}

\newcommand\pysr{\textit{PySR}\xspace}

\usepackage{amsmath}
\usepackage{amssymb}
\usepackage{mathtools}
\usepackage{amsthm}
\usepackage{physics}

\usepackage[capitalize,noabbrev]{cleveref}

\theoremstyle{plain}

\theoremstyle{definition}

\theoremstyle{remark}

\usepackage[textsize=tiny]{todonotes}

\begin{document}

\twocolumn[
\icmltitle{
Automated discovery \\
of interpretable gravitational-wave population models}

\begin{icmlauthorlist}
\icmlauthor{Kaze W.K Wong$^\ast$}{flatiron}
\icmlauthor{Miles Cranmer$^\ast$}{princeton}
\end{icmlauthorlist}

\icmlaffiliation{flatiron}{Center for Computational Astrophysics, Flatiron Institute, New York, NY 10010, USA}
\icmlaffiliation{princeton}{Department of Astrophysical Sciences, Princeton University, Princeton, New Jersey 08544, USA}

\icmlcorrespondingauthor{Kaze Wong}{kwong@flatironinstitute.org}

\icmlkeywords{Machine Learning, ICML}

\vskip 0.3in
]



\printAffiliationsAndNotice{\icmlEqualContribution} % otherwise use the standard text.

\begin{abstract}
We present an automatic approach to discover analytic population models for gravitational-wave (GW) events from data.
As more gravitational-wave (GW) events are detected, flexible models such as Gaussian Mixture Models have become more important in fitting the distribution of GW properties due to their expressivity.
However, flexible models come with many parameters that lack physical motivation, making interpreting the implication of these models challenging.
In this work, we demonstrate symbolic regression can complement flexible models by distilling the posterior predictive distribution of such flexible models into interpretable analytic expressions.
We recover common GW population models such as a power-law-plus-Gaussian, and find a new empirical population model which combines accuracy and simplicity.
This demonstrates a strategy to automatically discover interpretable population models in the ever-growing GW catalog, which can potentially be applied to other astrophysical phenomena.
\end{abstract}

\section{Introduction}
\begin{figure*}[hbt!]
\includegraphics[width=\textwidth]{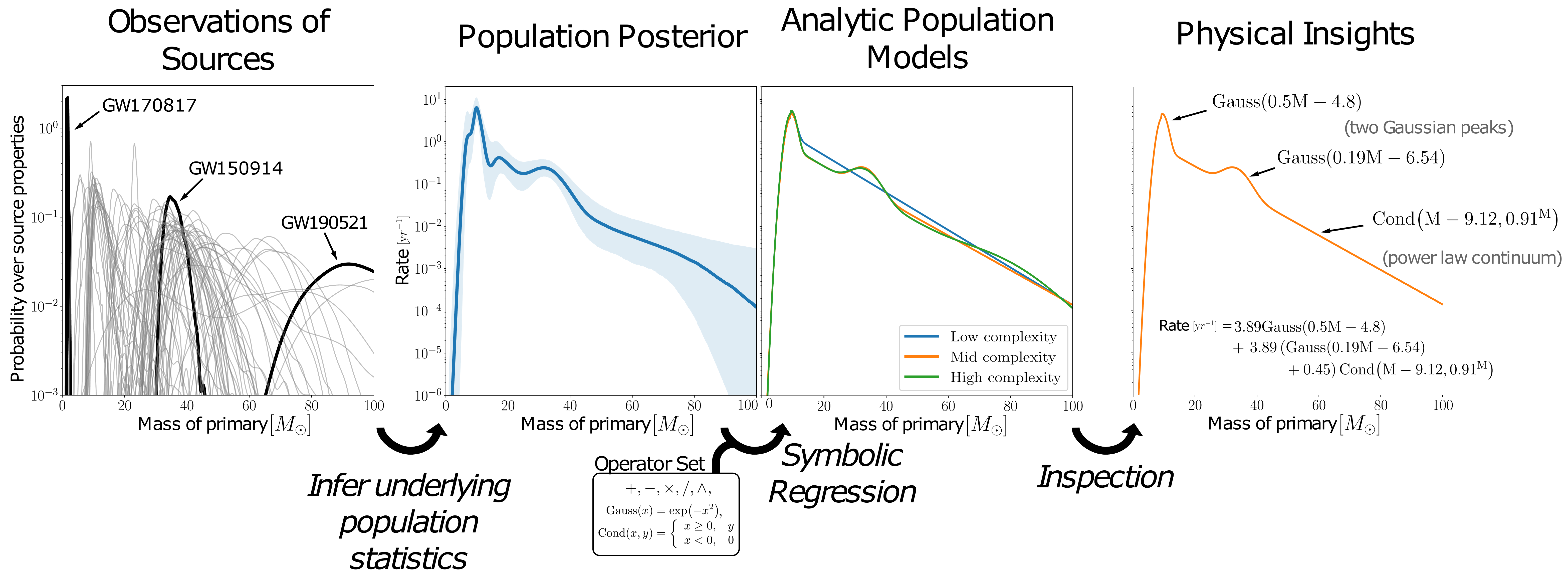}
\caption{An illustration of our proposed pipeline, which first
(\textit{far left})
requires observations of individual source properties (here, events shown from GWTC3);
then
(\textit{middle left}) 
infers the deconvolved population posterior density;
followed by (\textit{middle right})
using symbolic regression to search for an analytic model which approximates the density, using
operators common to existing population models; and, finally
(\textit{far right}) distilling physical insights
from the recovered population model.
}
%\caption{\textit{left}: posterior density of the primary mass from all events reported in GWTC3. Each line is the normalized posterior density for one event.
%\textit{middle}: posterior predictive distribution of the Gaussian mixture model reported in \cite{LIGOScientific:2021psn}.
%\textit{right}: equations tabulated in table \ref{tab:equations}}
\marginicon{\href{https://github.com/kazewong/SymbolicGWPopulation_paper/blob/main/src/scripts/figure1_subparts.py}{\faGithub}}
\label{fig:Gaussian_m1_fit}
\end{figure*}
The rate of gravitational wave (GW) detections is increasing exponentially as the sensitivity of GW detectors improves \cite{LIGOScientific:2021djp} since the discovery in 2015 \cite{Abbott:2016blz}.
The most update catalog of GW contains 79 events, the majority of them being Binary Black Holes (BBHs).
As more GW events are observed, more features in the distribution of GW events await to be discovered.
Currently, one of the most common ways to understand the observed GW population is through phenomenological modeling.
For example, to understand the distribution of primary mass in observed BBHs, one would hand-craft a model of the primary mass distribution, such as a power-law distribution \cite{LIGOScientific:2021psn}.
By comparing the model to the data, one can measure the parameters which characterize the model, such as the spectral index.

The main issue with this paradigm is the lack of scalability to a more complex dataset, since construction of sensible phenomenological models requires manual intervention.
To illustrate the shortcoming of this paradigm: 
when 11 events were announced in the first Gravitational-Wave Transient Catalog (GWTC1) in 2019 \cite{LIGOScientific:2018mvr}, a power law was sufficient to explain the primary mass distribution.
In GWTC2 (which is announced in 2020) \cite{LIGOScientific:2020kqk}, the community modified the simple power law to either a broken power law with two spectral indices, or with the addition of a Gaussian distribution.
Finally, the state-of-the-art model used in analyzing GWTC3 \cite{LIGOScientific:2021psn} further adds a spline on top of the power-law and Gaussian model to fit the residual \cite{Edelman:2021zkw}.
Since the complexity of the catalog grows with the size of the catalog,
it becomes increasingly difficult to develop an interpretable model that is flexible enough to explain the increasing rich set of features in the catalog . 

Alternatively, there exist studies using much more flexible models such as  Gaussian Mixture Models (GMMs) to analyze the catalog \cite{Tiwari:2020vym}.
While such a flexible model is powerful in fitting complex datasets,
it is difficult to interpret the physical meaning of the fitted parameters.
There stand two core reasons why such flexible models are hard to interpret:
first, these models often have many parameters;
in GWTC3, the GMM has over 100 parameters.
With so many degrees of freedom, it is non-trivial to analyze contributions from each mode and attempt to disentangle meaningful patterns.
Second, flexible models often employ bases that are not physically motivated.
For example, the GMM models has 11 independent Gaussian components in the mass-spin space.
While one or two Gaussians in mass or spin space alone can be physically motivated,
the additional components are included in the model fit for unexpected features.
Therefore, interpreting the importance of each component could be complicated by the components that are not physically motivated.

Symbolic regression (SR) is a machine learning technique which searches to space of analytic expressions to fit a dataset, which complements the weakness of a flexible model in understanding the distribution of GW.
However, since SR does not take advantage of using gradient information when searching for expressions,
and  SR searches a combinatorial space over expressions,
it is slow in comparison to traditional machine learning techniques such as Gaussian Mixture Models (GMMs).
Therefore, rather than fitting SR directly to the complex and expensive likelihoods found in GW event analysis, we first fit the GMM as a fast-to-evaluate approximation of the true posterior.
Then, we run SR to approximate the output of this GMM, which is feasible only because the loss function is then simply mean-squared error against the GMM, rather equal to an evaluation of the raw and expensive likelihood.
%While SR itself is not efficient enough to be used to analyze the data directly, using SR to interpret a flexible model that has already been fitted to the data can be very effective.

In this work, we apply SR to the mixture Gaussian model used to fit the primary mass distribution of GWTC3.
Specifically, we approximate the posterior predictive distribution produced of the Gaussian mixture model with SR.
This paper is structured as follows: we describe the GW data used in this work in \ref{sec:GWdata}.
We briefly review SR in section \ref{sec:method}.
We present our main result in section \ref{sec:result}.
We discuss the potential usage and extension of this work in section \ref{sec:discussion}.

\section{Method}
\label{sec:method}

\subsection{Symbolic regression}

While most ML algorithms optimize a fixed set number of real parameters,
symbolic regression is a machine learning (ML) algorithm which searches for analytic expressions which optimize some loss function.
The search typically employs a ``genetic algorithm,’’ which arranges symbols in a tree, evaluates the resultant expression, and repeats this process until a particular target is met.
Since models for physical phenomena are often specified in a language of concise mathematical expressions, symbolic regression can produce interpretable models compared to traditional ML, as the ``trained model'' generated by symbolic regression is in the same \textit{language} as existing physical models (i.e., analytic expressions).
Therefore, symbolic regression can produce insights for science in a way that traditional ML cannot.

\begin{table*}[hbt!]
    \begin{center}
    \begin{tabular}{ l l l l }
    \hline
    \hline
    Complexity &  Score & Loss & Expression for $\text{Rate}[\text{yr}^{-1}](M=M/M_\odot)$\\
    \hline
    \hline
    16 & 0.126   & 0.212  & $3.72 \rm{Cond}{\left(M - 9.27,0.9^{M} \right)} + 3.72 \rm{Gauss}{\left(0.52 M - 4.85 \right)}$\\[0.4em]
    % Add more space between rows of table^
    25 & 0.501   & 0.0352 & $3.89 \left(\rm{Gauss}{\left(0.19 M - 6.54 \right)} + 0.45\right) \rm{Cond}{\left(M - 9.12,0.91^{M} \right)}$\\
    & & & $ +\; 3.89 \rm{Gauss}{\left(0.5 M - 4.8 \right)}$\\[0.4em]
    48 & 0.00771 & 0.0108 & $\left(\rm{Cond}{\left(M - 9.42,0.62 \cdot 0.9^{M} \right)} + 1.44 \rm{Gauss}{\left(0.51 M - 4.88 \right)}\right)$\\
    & & & $\times\left(5.11 \rm{Gauss}{\left(0.06 M - 4.67 \right)} + 7.82 \rm{Gauss}{\left(0.17 M - 5.86 \right)} + 3.26\right)$\\
    \hline
    \hline
    \end{tabular}
    \caption{Expressions obtained through symbolic regression with \pysr.
    In the search we perform, there are 30 equations with different complexities.
    We select three representative equations from the Pareto front by setting three successive complexity ranges, and selecting the highest scoring expression in each range.}
    \label{tab:equations}
    \end{center}
\end{table*}

\paragraph{Algorithm.}
There are several different strategies for performing symbolic regression, all of them offering different advantages.
Some methods, such as \textit{SINDy} \cite{bruntonDiscoveringGoverningEquations2016}, are dictionary-based, and search for a sparse linear combination of hand-crafted terms.
Other methods, such as \textit{eureqa} \cite{schmidtDistillingFreeFormNatural2009}, are genetic algorithm-based, and, though slower, can find any expression that can be represented as a tree of predefined operators, variables, and constants.

We use the algorithm and software \pysr\footnote{\url{https://github.com/MilesCranmer/PySR}} \cite{cranmerPySRFastParallelized2020} for performing symbolic regression.
\pysr uses a genetic algorithm-style symbolic regression algorithm
\citep[see][for an early example]{kozaGeneticProgrammingMeans1994}.
Loosely inspired by \textit{eureqa},
\pysr searches the space of trees 
containing a basis set of specified operators, variables, and arbitrary real constants
to find expressions which fit relations in a given dataset.
\pysr uses ``regularized evolution'' for its search strategy 
\cite{realRegularizedEvolutionImage2019,realAutoMLZeroEvolvingMachine2020},
which itself is a variant on classic tournament selection
\cite{brindleGeneticAlgorithmsFunction1980,goldbergComparativeAnalysisSelection1991}.
This algorithm simulates natural selection: each expression is represented by a tree, corresponding to its ``genetic code.''
Expressions are subsampled, and the fittest expression in each group (using a metric which combines accuracy and simplicity) is used to breed new mutated expressions, as well as cross-bred with other top expressions.
\pysr adds several additional techniques onto these classic algorithms, such as explicit constant optimization using the \textit{BFGS} optimization algorithm \cite{fletcherPracticalMethodsOptimization1988}, and an adaptive complexity penalty, while allowing the use of custom operators and losses which is important for this work.

Internal to \pysr, a multi-objective optimization problem is solved: the evolution seeks to find expressions which are as accurate as possible, while being as simple as possible.
Formally, for our setting, \pysr solves the following optimization problem:
\begin{align}
    s_c &= 
        \mathop{\arg\min}_{
            \tiny
            \begin{array}{c}s\in\mathcal{S}\\ \abs{s} = c \end{array}
        }
        \mathcal{L}(s),
    \quad\text{for each } c\in \mathcal{C},
\end{align}
for the loss function $\mathcal{L}(s) = \sum_i \ell(s(\mathbf{x}_i), y_i),$
where $s$ is an expression, 
$\mathcal{S}$ is a space of possible expressions considered,
$\abs{s}$ is the complexity of an expression $s$,
$\ell$ is the loss function, and $\{(x_i, y_i)\}$ is the dataset
for $x_i$ each input and $y_i$ the label.
The most accurate expression at each $c$, denoted as $s_c$, is computed for a given set range of $c$, which we choose as $c\in\mathcal{C}\equiv \{1, \ldots, 50\}$.
Complexity is defined here
as the number of symbols in the expression (including repeats),
and $\ell$ as $\ell(\hat{y}_i, y_i) = \qty(\hat{y}_i - y_i)^2$.

Next, given the set of the best expressions at each complexity $\mathcal{S}^\ast = \{s_c\}_{c\in\{1,\ldots,80\}}$,
Next, we can construct the \textit{Pareto front}:
\begin{equation}
    \hat{\mathcal{S}}^\ast = \{s_c \ |\ \mathcal{L}(s_c) < \mathcal{L}(s_q) \forall q < c \},
\end{equation}
such that each remaining expression is more accurate than all simpler expressions.
Using this, we can define the \textit{score} of an expression as the slope of the Pareto front, equal to: $\text{score} = -\Delta(\log \mathcal{L}(s_c))/\Delta c,$
which is a heuristic for the quality of a particular equation, used in the symbolic regression literature \citep[see, e.g.,][]{schmidtDistillingFreeFormNatural2009,cranmerDiscoveringSymbolicModels2020}.

% \todo{Note how much computational power was used, what other PySR settings were used. Could link the script in the paper repo.}

\paragraph{Implementation for GWs}
% Describe reanalysizing the real data using the extracted expression
% After obtaining a number of 
Symbolic regression is an automated type of empirical model discovery---something which has driven the development of many theories in the physical sciences.
For example, Kepler's laws were an empirical model crafted by hand to fit patterns in the orbits of planets, and ultimately inspired Newton to discover a formula for gravitational force which could explain Kepler's laws \citep[see][for a review of this history]{hawkingShouldersGiantsGreat2004}.
Astronomy has seen increasing use of symbolic regression to find new empirical models in an automated way \citep[e.g.,][]{grahamMachineassistedDiscoveryRelationships2013,cranmerDiscoveringSymbolicModels2020,wadekarModelingAssemblyBias2020,delgadoModelingGalaxyhaloConnection2021,cranmerHistogramPoolingOperators2021,shaoFindingUniversalRelations2021}.

Here, we use symbolic regression for the first time to learn population density models in astrophysics.
One of the main advantages of symbolic regression is to learn a model into an interpretable language, and, therefore, we define a set of operators which frequently occur in GW population models: addition, subtraction, division, power laws, square, Gaussians (unary operator: $f(x)=\exp(-x^2)$), and conditional statements (binary operator: $g(x, y)=\operatorname{if}(x \geq 0, y, 0)$). 
To simplify our assumptions, all operators are assigned the same complexity of $1$.
We also introduce additional constraints on how operators may be nested inside each other to rule out uninterpretable expressions; these are defined in \cref{sec:constraints}.

\section{Gravitational-Wave Data}
\label{sec:GWdata}

GW data comes in the form of time series, in which each time sample carries very little physical meaning.
It is easier to understand the physical implication of a GW event through its inferred source properties \cite{Veitch:2014wba}.
In the case of a BBH, the source properties include the masses of the two merging black holes.
On top of extracting physics from individual events,
we can also learn about astrophysical processes through understanding of the distribution of source properties, such as the mass distribution of black holes \cite{2019PASA...36...10T,Vitale:2020aaz}.

In principle, one can apply symbolic regression to fit these distributions directly.
Unfortunately, GW observations are extremely noisy. As a result, all inferred properties about the source such as the observed masses or distance to the source come with sizable uncertainty, often comparable to the prior used in inferring their source properties.
This also means the posterior distribution of these inferred properties often has a non-trivial shape that the typical Gaussian approximation of ``error bar’’ does not hold.
Moreover, since there are only 79 announced events, the shot noise makes it challenging to extract meaningful information if we first stack the events and fit the distribution with symbolic regression.

In order to maximally extract information from the GW catalog, the community has employed Hierarchical Bayesian Analysis (HBA) to analyze the distribution of GW events.
HBA takes advantage of the fact that measurement uncertainties for each event are conditionally independent from the population model, so we can deconvolve the measurement uncertainty for each event (\citealt{10.1214/10-AOAS439} describes this as extreme deconvolution) while fitting for the underlying distribution of GW events\footnote{For interested readers, \cite{Mandel:2018mve,Vitale:2020aaz,2019MNRAS.484.4008G} are excellent references with practical examples on how to apply the HBA framework to GW data.
}.

\newcommand\ppd{posterior predictive distribution\xspace} %PPD\xspace
\newcommand{\defineppd}{} % (\ppd)

One product coming out of this HBA pipeline is the posterior predictive distribution \defineppd for the properties of interest, i.e., a set of probability density functions that encapsulate the inferred source distribution and the uncertainty on the population level.
It is much easier to find an interpretable form of the source distribution by fitting the \ppd with SR.
In this study, we focus on understanding the distribution of binary black hole primary mass, i.e., the heavier mass in the binary.
One of the most flexible models that is fitted to this set of data is a Gaussian mixture model \cite{Tiwari:2020vym}.
To generate the data that we feed to the SR pipeline,
we select many evaluation points in the primary mass axis,
then we take the median of the \ppd given by the Gaussian mixture model as the value we are trying to predict.
We also estimate the uncertainty of the value at each evaluation point using the $68\%$ confidence interval.

\section{Result}
\label{sec:result}

We show the workflow of this work in figure \ref{fig:Gaussian_m1_fit}.
On the leftmost panel, we show the posterior density in primary mass for every event in GWTC3.
In the left-middle panel, we show a GMM fitted to the data given by \cite{Tiwari:2020vym}.
Given the \ppd shown in the middle panel, we use symbolic regression to extract effective descriptions of the data. 
As we increase the complexity of the equation, it is natural to expect that the equation should fit the data better than a lower complexity equation since it has additional free parameters.
At the same time, an equation with higher complexity could either be overfitting or difficult to interpret.
\pysr uses a score (see \citealt{cranmerDiscoveringSymbolicModels2020} for details) as a metric normalized to complexity to show how well an equation is fitting the data.
At each complexity, \pysr outputs the equation with the lowest loss.
To select the equations shown in figure \ref{fig:Gaussian_m1_fit}, we first make an accuracy cut based on the loss of the function such that all expressions have a loss lower than $0.3$, so the selected equation would still be a good fit to the data.
We then consider the expressions with the highest score among three ranges: $c\in [1, 20]$,
$c\in[20, 35]$, and $c\in[35, 50]$.
These ranges are set to give us a variety of complexity expressions to study.
The equations selected under these criteria are tabulated in table \ref{tab:equations}.

The lowest complexity (16) equation returned is a decaying exponential function with a Gaussian bump at the lower end of the mass distribution.
This agrees with a recent study pointing out there is a very strong excess of events at $m_{1}\sim 10\ M_{\odot}$,
which corresponds to a smooth turn-on of the black hole mass spectrum \cite{Talbot:2018cva}.
In order to account for the bump around $30\ M_{\odot}$,
the exponential part of this equation has a much higher normalization compared to the others.
This agrees with the analysis done in GWTC3, where a single power law seems to have trouble accounting for the excess of events around $m_{1}\sim 30\ M_{\odot}$\cite{LIGOScientific:2021psn}.

Compared to the lowest complexity equation, the middle complexity (25) equation instead prefers a lower normalization of the exponential function,
but account for the bump around $30\ M_{\odot}$ with a Gaussian component,
which is proposed to be driven by a pile-up from pulsational pair-instability supernovae \cite{Talbot:2018cva}.
This equation is similar to one of the simplest phenomenological models fitted to GWTC3, which is a power law mixed with a Gaussian bump.
The difference between this equation and the one presented in GWTC3 is this equation has an additional bump at $m_{1}\sim 10\ M_{\odot}$,
which is also present in the lower complexity equation.
This suggests the first bump at lower mass has more statistical significance than the second bump.
Although the middle complexity equation is more complex, the loss is about one order of magnitude lower than the next best one with lower complexity (16).
As a result, the score of this equation is higher than the lower complexity equation, making it a favorable equation.

At the highest complexity (48), one more Gaussian component is added to the equation account for the tapering at the higher mass end. 
Interestingly, the new component added in this equation does not try to fit the wiggles in between the first bump and the second bump.
The significance of any features other than the two bumps and the continuum is still an open question.
Our result hints the wiggles between the bumps seem to be less significant than the extra tapering of the continuum.
In addition, none of the equations prefers to fit the extra hump before the $10\ M_{\odot}$ peak.
This could mean the little hump could be an artifact in the Gaussian mixture model instead of a truly statistically significant feature.
However, the loss of this equation is similar to the middle complexity (25) equation, at the same time being almost twice as complex. As a result, the score is much lower, which means the tapering is not necessarily needed.

\section{Discussion}
\label{sec:discussion}

% Brief summary of the paper
In this paper, we use symbolic regression to extract interpretable equations from the \ppd of a flexible model fitted to the catalog of GWs.
We show SR can discover equations which are similar to state-of-the-art phenomenological models.
On top of finding forms that have already been discovered,
our result also shed insights on an open question---whether there exist additional features other than the two prominent peaks in the primary mass distribution \cite{Tiwari:2021yvr}.
Since the highest complexity equation does not significantly improve the fitting accuracy, one should take caution when interpreting the tapering of the continuum or wiggles between the two peaks as their significance are unclear.

The main message of this work is that by combining flexible models with symbolic regression, we can discover interpretable equations that can describe the increasingly complex GW catalog.
Compared to traditional methods such as phenomenological modeling, which relies on an individual scientist's insight,
our pipeline requires minimal human intervention.
 This technique is not a model selection method by itself, but a pipeline to discover models for downstream tasks such as model selection and parameter estimation.
Once we obtain a set of equations, we can insert the expressions to a standard HBA pipeline to estimate the parameters of the model and select between models.
This allows the community to discover interpretable patterns in the GW catalog in a data-driven instead of hand-crafted model-driven manner.
In this sense, our method is more scalable to more complex and higher-dimensional datasets.

In this work, we focus on analyzing only the primary mass distribution.
Our method is trivially generalizable to higher-dimensional data, such as the joint space of both the primary mass and the secondary mass.
So far there are relatively few phenomenological models which can capture correlations between observables,
mainly due to the difficulty of manually constructing reasonable models to account for these correlations.
At the same time, the most advanced flexible model in the literature is an n-dimensional histogram, which is hard to interpret.
Our method provides a natural way to explore and to explain these correlations.

Although we focus on GW in this work, this method can be easily extended to other population analysis problems.
For example, given the recently released Gaia DR3 dataset \cite{Collaboration2016TheGM}, it might be possible to apply this method to understanding various subpopulations of  stars in the Milky Way.
As size of datasets grows larger in astronomy, this work provides a simple and scalable pathway to uncover trends and relations in astrophysical catalogs.

\bibliographystyle{icml2022}
\bibliography{bib}

\clearpage
\appendix
\section{Operator Constraints}
\label{sec:constraints}

With full flexibility over how symbolic operators can be arranged up to a maximum number, surprisingly many combinations exist which are simple and accurate, yet completely uninterpretable.
Typically, a genetic algorithm might find that a particular deeply nested combination of operators might perform very well, with the overall complexity being low.
However, we wish to find a model that is interpretable, and so we provide some additional constraints here.

First, we implement limits on which operators can be nested, and how deeply.
We impose the following nested constraints:
\begin{enumerate}
    \item No operator with arity of 1 (i.e., an operator taking a single argument) may be nested with another arity-1 operator, unless the operator is square.
    \begin{enumerate}
        \item However, we permit a square to be nested inside itself once, as such a combination may be used to make a quadratic operation with few operations.
    \end{enumerate}
    \item The division operator may only be nested inside itself once.
    \item The power operator may not contain any arity-1 operator.
\end{enumerate}
We also impose the following constraints on maximum complexity within each operator:
\begin{enumerate}
    \item The power operator may only have a maximum complexity of 9 in its base, and 3 in its exponent.
    \item The division operator may have any complexity in the numerator, but a maximum complexity of 7 in the denominator.
    \item The Gauss operator may only have a maximum complexity of 11 in its argument.
    \item The conditional operator may only have a maximum complexity of 5 in its condition, and 11 in its resultant argument.
\end{enumerate}
These prevent \pysr from finding expressions which, despite being low-complexity and high-accuracy, are still uninterpretable to a human scientist due to unusual combinations of operators compared to typical expressions found throughout physics and astrophysics.

\end{document}